\documentclass[twocolumn,showpacs,preprintnumbers,amsmath,amssymb]{revtex4}
\usepackage{graphicx}
\usepackage{dcolumn}
\usepackage{color}
\usepackage{bm}
\begin{document}
\title {Dissipative Diamagnetism --- A Case Study for Equilibrium and Nonequilibrium Statistical Mechanics of Mesoscopic Systems.}
\vskip -0.5cm
\author{ Malay Bandyopadhyay and  Sushanta Dattagupta}
\affiliation{Nano Science Unit, S.N. Bose National Centre for Basic Sciences,JD Block, Sector III, Salt Lake City, Kolkata 700098, India.}
\date{\today}
\begin{abstract}
Using the path integral approach to equilibrium statistical physics the effect of dissipation on Landau diamagnetism is calculated. The calculation clarifies the essential role of the boundary of the container in which the electrons move. Further, the derived result for diamagnetization also matches with the expression obtained from a time-dependent quantum Langevin equation in the asymptotic limit, provided a certain order is maintained in taking limits. This identification then unifies equilibrium and nonequilibrium statistical physics for a phenomenon like diamagnetism, which is inherently quantum and strongly dependent on boundary effects. 
\end{abstract}
\maketitle
An unconventional approach to statistical physics, which may be referred to as the Einstein approach, involves the derivation of equilibrium results  from the long-time limit of time-dependent equations \cite{kadanoff}. Specifically, a set of Langevin equations (or their equivalent in the phase space, called the Fokker-Planck equation), with built-in detailed balance conditions, can naturally yield asymptotic results that can be independently calculated from the Gibbs ensemble idea of statistical physics. The underlying concept is physically appealing because not only does it sidetrack the issue of ergodicity, which is assumed at the outset in the Gibbs prescription, it also connects directly to experimental measurements, which necessarily involve time-averages. In this centenary of Einstein's {\em annus mirabilis} it is momentously appropriate to assess the validity and usefulness of this  approach to statistical physics, that relies on the central paradigm of Brownian motion \cite{brown}.\\
\noindent
Given this motivation we want to further explore the Einstein approach in this  Letter by going beyond the classical into the quantum domain. The phenomenon of interest happens to be intrinsically and essentially quantum mechanical --- it relates to the issue of diamagnetism exhibited by a collection of electrons subjected to an applied magnetic field. Diamagnetism is an enigmatic subject in that not only does it require a quantum treatment, as provided by the landmark work of Landau \cite{landau}, but it also needs a careful analysis of the boundary of the container in which the electrons are constrained to move. As has been discussed lucidly by Van Vleck \cite{van vleck}, the boundary electrons exactly cancel the contribution of the bulk electrons, in classical  physics, leading to the celebrated Bohr-Van Leeuwen theorem \cite{bohr}. However this cancellation is incomplete in the quantum regime, because as Peierls points out \cite{peirls}, it is the boundary electrons which have the "skipping orbits" that yield the edge currents, familiar also in quantum Hall effect \cite{datta}, which make an essential contribution to diamagnetism. A few years ago, we have examined the question of Landau diamagnetism in a dissipative and confined system \cite{sdg}.\\
\noindent
The following issues were addressed in I: (a) the approach to equilibrium of a quantum dissipative system, the analysis of which brings out the subtle role of boundary electrons, (b) the effect of dissipation on Landau diamagnetism, an equilibrium property, (c) quantum - classical crossover as the system transits from the Landau to the Bohr-Van Leeuwen regime as a function of damping and (d) the combined effect of dissipation and confinement on Landau diamagnetism, the latter arising from coherent cyclotron motion of the electrons. The item (d) is particularly relevant in the context of intrinsic decoherence in mesoscopic structures in view of heat bath induced influence \cite{datta,mohanty,imry}. Dissipation was incorporated in I with the aid of a quantum Langevin equation, driven by a systematic Lorentz force, that can be derived from an underlying Hamiltonian in a system-plus-bath formulation in which the bath degrees of freedom are integrated out \cite{ford}. In the infinite past the bath is assumed to be in thermal equilibrium such that the fluctuations of its degrees of freedom are governed by quantum statistics. Thus, detailed balance conditions are automatically expressed through a `fluctuation-dissipation' relation that relates the noise spectrum to the damping term in the quantum Langevin equation.\\
\noindent
The starting point of I as indeed in this Letter is the Feynman-Vernon \cite{feynman} Hamiltonian for a charged particle $e$ in a magnetic field $\vec{B}$:
\begin{eqnarray}
{\cal{H}} & = & \frac{1}{2m}\omega_{0}^{2}{\vec{x}}^{2}+ \frac{1}{2m}\Big(\vec{p} - \frac{e \vec{A}}{c}{\Big)}^{2} \nonumber \\
& & + \sum_{j=1}^{N}{\Big[}\frac{1}{2m_{j}}\vec{p_{j}}^{2} +\frac{1}{2}m_{j}{\omega_{j}}^{2}({\vec{x}}_{j}-\vec{x})^{2}{\Big]},
\end{eqnarray}
where the first term is the Darwin \cite{darwin} term representing a confining potential, $\vec{p}$ and $\vec{x}$ are the momentum and position operators of the particle, ${\vec{p}}_{j}$ and ${\vec{x}}_{j}$ are the corresponding variables for the bath particles, and $\vec{A}$ is the vector potential. The bilinear coupling between $\vec{x}$ and $\vec{x}_{j}$ as envisaged in Eq. (1) has been the hall mark of the Caldeira-Leggett approach to dissipative quantum mechanics \cite{legget1,legget2}. Assuming the $\vec{B}$ field to be along the $z$-axis, all the vectors in Eq. (1) can be taken to lie in the $xy$-plane. From the quantum Langevin equation , derived from Eq. (1) by following the steps mentioned above, the nonequilibrium or time-dependent magnetization along the $z$-axis, $M_{z}(t)$ is computed in I. It is important to note that the Landau answer for the magnetization, in equilibrium, ensues from $M_{z}(t)$ only by following the limiting procedures in a specific order, viz; by first taking t $\rightarrow \infty$ and then setting $\omega_{0} \rightarrow$ 0. If these two limits are interchanged one ends up with a piece of the Landau answer that misses out the boundary contribution. \\
\noindent
Having laid down the background to the myriad perplexing issues concerning diamagnetism we pose and answer the following question in this Letter. Should we not be able to calculate the equilibrium magnetization directly from Eq. (1) by following the usual Gibbsian statistical mechanics in which all the terms in Eq. (1) are treated on the same footing and there is no separation between what is a system and what is a bath? If the answer to this question is in the affirmative and the resultant magnetization matches with the result derived in I in the `equilibrium limit' that would indeed lend the Einstein approach yet another foundational basis.\\
\noindent
The energy eigenvalues for the Hamiltonian in Eq. (1) have been computed by Hong and Wheatley \cite{hong}. However our method of calculation is based on the functional integral approach to statistical mechanics which we find to be the most convenient tool for studying charged particle dynamics in a magnetic field \cite{feynman1,feynman2,kleinert,weiss,ingold}. The Euclidean action corresponding to the Hamiltonian in Eq. (1) can be written as :

\begin{equation}
{\cal{A}}_{e} = \int_{0}^{\hbar \beta}d\tau [{\cal{L}}_{S}(\tau) + {\cal{L}}_{B}(\tau) + {\cal{L}}_{I}(\tau)],
\end{equation} 
where the subscripts S, B and I stand for `system', `bath' and `interaction' respectively. The corresponding Lagrangians are enumerated as:
\begin{equation}
{\cal{L}}_{S}(\tau) = \frac{M}{2}\Big[\dot {\vec{x}}(\tau)^{2} + \omega_{0}^{2}\vec{x}(\tau)^{2} - \omega_{c}(\vec{x}(\tau) \times \dot{\vec{x}}(\tau))_{z}\Big],
\end{equation}
where $\omega_{c}= \frac{eB}{Mc}$, is the cyclotron frequency,
\begin{equation}
{\cal{L}}_{B}(\tau) = \sum_{j=1}^{N} \frac{1}{2}m_{j}[\dot {\vec{x}_{j}}(\tau)^{2} + \omega_{j}^{2}\vec{x}_{j}(\tau)^{2}],
\end{equation}
\begin{equation}
{\cal{L}}_{I}(\tau) = \sum_{j=1}^{N} \frac{1}{2}m_{j}\omega_{j}^{2}[\vec{x}(\tau)^{2} - 2\vec{x}_{j}(\tau)\cdot\vec{x}(\tau)] .
\end{equation}
We introduce now imaginary time Fourier series expansion of system variables and bath variables as follows:
\begin{eqnarray}
\vec{x}(\tau) & = & \sum_{n}\vec{\tilde{x}}(\nu_{n})e^{-i\nu_{n}\tau}, \\
\vec{x}_{j}(\tau) & = & \sum_{n}{\vec{\tilde{x}}}_{j}(\nu_{n})e^{-i\nu_{n}\tau},
\end{eqnarray}
where the  Bosonic Matsubara frequencies $\nu_{n}$ are given by
\begin{equation}
\nu_{n} = \frac{2\pi n}{\hbar \beta}, \quad {n} = 0,\pm 1,\pm 2, .....,
\end{equation}
The system-part of the action in terms of Fourier components is:
\begin{eqnarray}
{\cal{A}}_{e}^{S} & = & \frac{M}{2}\hbar \beta \sum_{n}\Big[(\nu_{n}^{2} + \omega_{0}^{2})(\vec{\tilde{x}}(\nu_{n})\cdot \vec{\tilde{x^{*}}}(\nu_{n})) \nonumber \\
&  & +\omega_{c} \nu_{n}(\vec{\tilde{x}}(\nu_{n}) \times \vec{\tilde{x^{*}}}(\nu_{n}))\Big]. 
\end{eqnarray}
In deriving Eq. (9) we have used the identity:
\begin{equation}
\int_{0}^{\hbar \beta} d\tau e^{-i\tau(\nu_{n}+\nu_{n^{\prime}})} = \hbar \beta \delta (n +n^{\prime}).
\end{equation}
Following the detailed treatment given by Weiss \cite{weiss}, the combined contributions of the bath and the interaction terms to the action can be written as:
\begin{equation}
{\cal{A}}_{e}^{B-I} = \frac{M}{2}\hbar \beta \sum_{n}\xi(\nu_{n})(\vec{\tilde{x}}(\nu_{n})\cdot\vec{\tilde{x}^{*}}(\nu_{n})) , 
\end{equation}
where 
\begin{equation}
 \xi(\nu_{n}) = \frac{1}{M} \sum_{j=1}^{N} m_{j}\omega_{j}^{2}\frac{\nu_{n}^{2}}{(\nu_{n}^{2}+\omega_{j}^{2})}.
\end{equation}
Introducing the spectral density for bath excitations as:
\begin{equation}
J(\omega) = \frac{\pi}{2}\sum_{j=1}^{N}m_{j}\omega_{j}^{3}\delta(\omega - \omega_{j}), 
\end{equation}
we may rewrite
\begin{equation}
\xi(\nu_{n}) = \frac {2}{M\pi}\int_{0}^{\infty}d\omega \frac{J(\omega)}{\omega} \frac{\nu_{n}^{2}}{(\nu_{n}^{2} + \omega^{2})}.
\end{equation}
Now combining Eq. (11) with Eq. (9), the full action can be expressed as:
\begin{eqnarray}
{\cal{A}}_{e} & = & \frac{M}{2}\hbar \beta \sum_{n}[(\nu_{n}^{2} + \omega_{0}^{2} + \nu_{n}\tilde{\gamma}(\nu_{n}))(\vec{\tilde{x}}(\nu_{n})\cdot \vec{\tilde{x^{*}}}(\nu_{n}))  \nonumber \\
&  & +\omega_{c} \nu_{n}(\vec{\tilde{x}}(\nu_{n}) \times \vec{\tilde{x^{*}}}(\nu_{n}))], 
\end{eqnarray}
where the `memory-friction' is given by
\begin{equation}
\tilde{\gamma}(\nu_{n}) =\frac{2}{M\pi}\int_{0}^{\infty}d\omega \frac{J(\omega)}{\omega} \frac{\nu_{n}}{(\nu_{n}^{2} + \omega^{2})}.
\end{equation}
Note that $\vec{\tilde{x}}(\nu_{n})$ is a two-dimensional vector ($\tilde{x}(\nu_{n}),\tilde{y}(\nu_{n})$). Introducing then normal modes:
\begin{eqnarray}
\tilde{z}_{+}(\nu_{n}) & = & \frac{1}{\sqrt2}(\tilde{x}(\nu_{n})+i\tilde{y}(\nu_{n})) \nonumber \\
\tilde{z}_{-}(\nu_{n}) & = & \frac{1}{\sqrt2}(\tilde{x}(\nu_{n})-i\tilde{y}(\nu_{n})), 
\end{eqnarray}
Eq. (15) can be rewritten in a `separable' form:
\begin{eqnarray}
{\cal{A}}_{e} & = &\frac{M}{2}\hbar \beta \sum_{n}\Big[(\nu_{n}^{2} + \omega_{0}^{2} + \nu_{n}\tilde{\gamma}(\nu_{n})+i\omega_{c}\nu_{n}) \nonumber \\
& &(\tilde{z}_{+}(\nu_{n})\tilde{z}^{*}_{+}(\nu_{n})) \nonumber \\
& & + (\nu_{n}^{2} + \omega_{0}^{2} + \nu_{n}\tilde{\gamma}(\nu_{n})-i\omega_{c}\nu_{n}) \nonumber \\
& & (\tilde{z}_{-}(\nu_{n})\tilde{z}^{*}_{-}(\nu_{n}))\Big]. 
\end{eqnarray}
The partition function is then given by:
\begin{eqnarray}
{\cal{Z}} & = & {\prod}_{n}\int d\tilde{z}_{+}(\nu_{n})d\tilde{z}^{*}_{+}(\nu_{n})d\tilde{z}_{-}(\nu_{n})d\tilde{z}^{*}_{-}(\nu_{n}) \nonumber \\
& & \exp\Big[-\frac{1}{2}M\beta(\nu_{n}^{2}+ \omega_{0}^{2} + \nu_{n}\tilde{\gamma}(\nu_{n})+i\omega_{c}\nu_{n}) \nonumber \\
& & (\tilde{z}_{+}(\nu_{n})\tilde{z}^{*}_{+}(\nu_{n}))\Big]\nonumber \\
& & \exp\Big[-\frac{1}{2}M\beta(\nu_{n}^{2}+ \omega_{0}^{2} + \nu_{n}\tilde{\gamma}(\nu_{n})-i\omega_{c}\nu_{n}) \nonumber \\
& & (\tilde{z}_{-}(\nu_{n})\tilde{z}^{*}_{-}(\nu_{n}))\Big]\nonumber \\
& = & \frac{2\pi}{M\beta}{\prod}_{n}\Big[(\nu_{n}^{2}+ \omega_{0}^{2} + \nu_{n}\tilde{\gamma}(\nu_{n}))^{2}+ \omega_{c}^{2}\nu_{n}^{2}\Big]^{-1}. \nonumber \\ 
\end{eqnarray}
In view of Eqs. (8) and (16) the Helmholtz Free energy $\cal{F}$ can be deduced from Eq. (19) as   
\begin{eqnarray}
{\cal{F}} & = & \frac{1}{\beta}\ln\Big(\frac{M\beta\omega_{0}^{4}}{2\pi}\Big)  \nonumber \\
&  &  + \frac{2}{\beta}\sum_{n=1}^{\infty}\ln\Big[(\nu_{n}^{2}+ \omega_{0}^{2} + \nu_{n}\tilde{\gamma}(\nu_{n}))^{2} + \omega_{c}^{2}\nu_{n}^{2}\Big], \nonumber \\ 
\end{eqnarray}
where the first term is independent of the magnetic field and owes its existence purely due to the Darwinian constraining potential. Equation (20) contains all the thermodynamic properties, the most important of which is the {\em magnetization} given by the negative derivative of ${\cal{F}}$ with respect to $B$ :
\begin{eqnarray}
{\cal{M}} & = & -\sum_{n=1}^{\infty}\frac{\frac{4}{\beta B}\omega_{c}^{2}\nu_{n}^{2}}{[(\nu_{n}^{2}+ \omega_{0}^{2} + \nu_{n}\tilde{\gamma}(\nu_{n}))^{2} + \omega_{c}^{2}\nu_{n}^{2}]},
\end{eqnarray}
 Equation (21) identically matches with the asymptotic ($t\rightarrow \infty $ ) limit of the expression obtained by Li {\em etal} \cite{li} from a quantum Langevin equation formulation. Further, in the so-called ohmic dissipation model for which \cite{legget2}
\begin{equation}
J(\omega) = M\gamma \omega,
\end{equation}
the expression (21), upon using the identity:
\begin{equation}
\coth(z) = \frac{1}{z} + \sum_{n=1}^{\infty}\frac{2z}{(z^{2}+n^{2}\pi^{2})},
\end{equation}
also yields the asymptotic result of I, for $\omega_{0}=0$ (cf. Eq. (19) of I). The ohmic case is relevant for electron-hole excitations in a Fermionic bath whereas the non-ohmic case applies to a phononic heatbath \cite{weiss}. \\
\noindent
Equation (21) embodies several tantalizing results which deserve special comments: (1) The diamagnetization is one of the rare equilibrium properties which depends directly on the damping parameter $\gamma$. Seldom is dissipation discussed in text books within the realm of what we call equilibrium statistical mechanics, based on the Gibbs ensemble. The fact that $\gamma$ is a measure of dissipation has been amply demonstrated in I, wherein we had shown how by increasing $\gamma$, ${\cal{M}}$ changes from the Landau to the Bohr-Van Leeuwen expressions --- an example of coherence-to-decoherence transition in an open quantum system \cite{sdg1}. (2) Diamagnetism as a material property is seen to be situated at the crossroads of thermodynamics and transport phenomena. The thermodynamic nature of the property is rooted on its being able to be calculated from the free energy, as shown here. On the other hand, diamagnetism, like the Drude conductivity \cite{ashcroft}, is also based on transport mechanism in that it is related to the expectation value of the operator $(\vec{r} \times \vec{v})$ (see I). Because the velocity $\vec{v}$ appears explicitly, dissipative diamagnetism naturally connects to the fundamental frictional material property, viz. resistance, in view of the fact that $\gamma ^{-1}$ is related to the Drude relaxation time \cite {sdg2}. Again we are not aware of any other phenomenon that lies at the juxtaposition of thermodynamics, which is derived from a partition function and transport, that is usually treated in kinetic theory. (3) Normally, in statistical mechanics, a thermodynamic limit is taken as a result of which surface contributions to bulk become irrelevant. However, for diamagnetism the surface enters crucially, as argued above; even though, there are fewer surface electrons than in the bulk, their contribution to the operator $\vec{r}$ in $(\vec{r} \times \vec{v})$  is substantial. A remarkable feature of diamagnetism is the need to first calculate the magnetization in the thermodynamic limit and then switch the boundary off i.e. by setting $\omega_{0}=0$. One related issue is the environment  induced dissipation which happens to be a ubiquitous attribute of a mesoscopic system. Additionally, because for a mesoscopic system surface effects are non-negligible, the present study has a bearing on our understanding of mesoscopic structures. While points (1), (2) and (3) connote to thermal equilibrium we want to now make a few remarks on the significance of our results for the approach-to-equilibrium, in the present context: (4) usually this question is discussed in a system-plus-bath approach, within a master equation for the density operator. The subject of quantum optics is replete with such approaches wherein the interaction between the system and the bath is assumed weak and is consequently treated in the socalled Born-Markov approximation \cite{agarwal}. The result is, although the approach to equilibrium does depend on relaxation parameters such as damping the equilibrium results themselves are independent of such parameters. Thus the density operator approaches a Boltzmann distribution characterized by the Hamiltonian for the system alone. In contrast, the presently derived dissipative diamagnetization, which can also be computed from the nonequilibrium method of I, does depend explicitly on damping, as has been also emphasized under point (1) above. The reason is, like in the much studied problem of quantum dissipation of a harmonic oscillator \cite{grabert}, the system-bath coupling is so strong that it needs an exact treatment. Thus the degrees of freedom of the entire many body system are inexorably entangled with each other and therefore, it is no longer meaningful to separate what is a system from what is a bath. (5) Finally, a related point to (4) is in connection with the essential quantum nature of diamagnetism. As has been argued by Jayannavar and Kumar \cite{kumar}, not only is there no classical diamagnetism --- due to the Bohr-Van Leeuwen theorem --- there is no dissipative classical diamagnetism either. Thus, the nonequilibrium, classical diamagnetization relaxes to {\em zero}, a damping-independent result. The same is true for the classical damped harmonic oscillator. In that case the time-dependent probability distribution for the underlying Ornstein-Uhlenbeck-process \cite{oz} relaxes to the equilibrium Boltzmann distribution, free of damping, even though the system-bath coupling is treated exactly through the classical Langevin equations \cite{zwanzig}. Therefore, we emphasize once again that the appearance of damping terms in equilibrium answers, as discussed under points (4) and (1), is an intrinsically non-classical aspect.         
{\section*{Acknowledgement}}
We thank Sansaptak Dasgupta and Prosenjit Dutta for discussion, and B. M. Deb, B. Dutta Roy and J. Garcia-Palazios for their critical reading of the manuscript.\\

\end{document}